\documentclass[english,aps,preprint,superscriptaddress, prl,english,amsmath,amssymb]{revtex4-2}
\usepackage{lmodern}
\usepackage[T1]{fontenc}
\usepackage[latin9]{inputenc}
\usepackage{amsmath}
\usepackage{amsthm}
\usepackage{amssymb}
\usepackage{graphicx}

\makeatletter
\usepackage[T1]{fontenc}
\usepackage{amsmath}
\usepackage{amsthm}
\usepackage{amssymb}
\usepackage{graphicx}

\usepackage{graphicx}
\usepackage{float}
\usepackage{xcolor}
\usepackage[hypertexnames=false]{hyperref}
\hypersetup{
	breaklinks = true,
    colorlinks = true,
    citecolor = {blue},
	urlcolor = {blue},
	linkcolor = {blue}
}

\makeatother

\usepackage{babel}
\begin{document}
\title{Hidden free energy released by explicit parity-time-symmetry breaking}
\author{Hong Qin}
\email{hongqin@princeton.edu }

\affiliation{Princeton Plasma Physics Laboratory, Princeton University, Princeton,
NJ 08540, USA}
\affiliation{Department of Astrophysical Sciences, Princeton University, Princeton,
NJ 08540, USA}
\author{William Dorland}
\email{dorland@pppl.gov}

\affiliation{Princeton Plasma Physics Laboratory, Princeton University, Princeton,
NJ 08540, USA}
\affiliation{Department of Physics, University of Maryland, College Park, MD 20742,
USA}
\author{Ben Y. Israeli}
\email{ben.israeli@weizmann.ac.il}

\affiliation{Department of Physics of Complex Systems, Weizmann Institute of Science,
Rehovot, Israel}
\begin{abstract}
It is shown that the familiar two-stream instability is the result
of spontaneous parity-time (PT)-symmetry breaking in a conservative
system, and more importantly, explicit PT-symmetry breaking by viscosity
can destabilize the system in certain parameter regimes that are stable
when viscosity vanishes. This reveals that complex systems may possess
hidden free energies protected by PT-symmetry and viscosity, albeit
dissipative, can expose the systems to these freed energies by breaking
PT-symmetry explicitly. Such a process is accompanied by instability
and total variation growth.
\end{abstract}
\maketitle
\begin{figure}[ht]
\centering \includegraphics[width=12cm]{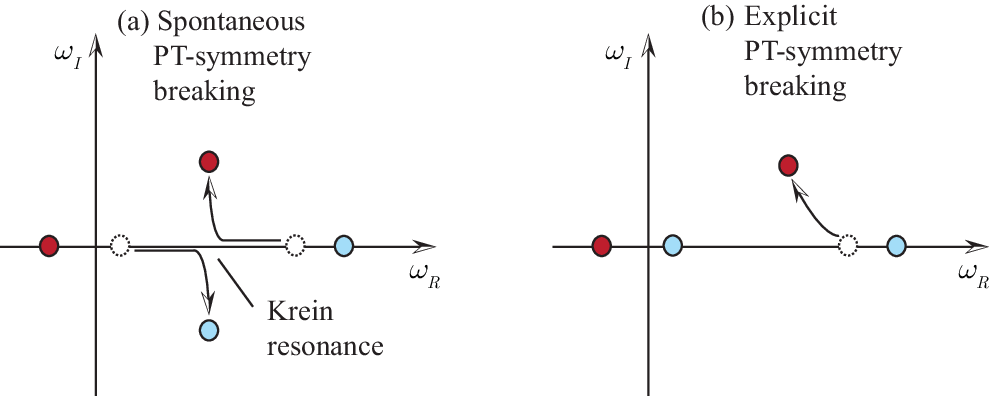} \caption{Spontaneous (a) and explicit (b) PT-symmetry breaking for a 4D system.
Red and blue circles indicate positive- and negative-action modes,
respectively.\textbf{ }The route of destabilization illustrated in
(b) is forbidden by PT symmetry. }
\label{fig:1}
\end{figure}

One important origin of the complexity associated with plasma dynamics
is the interaction between different charged components in plasmas.
The two-stream interaction between two charged components mediated
by electromagnetic or electrostatic field have been identified and
studied extensively for many applications and devices \citep{Kueny1995,Qin2000BEST,Davidson01-all,Zimmermann2004,LashmoreDavies2007,Morrison2014,Qin2014TwoStream,Zhang2016GH,Israeli2023},
including the classical Kelvin-Helmholtz instability and bump-on-tail
instability as special cases. It is also used as a standard benchmark
for numerical algorithms and theoretical methods. In the present study,
we show that conservative two-stream interaction is parity-time (PT)
symmetric and the familiar two-stream instability is driven by the
spontaneous PT-symmetry breaking (see Fig.\,\ref{fig:1}a). More
importantly, we show that finite viscosity breaks PT-symmetry explicitly
(see Fig.\,\ref{fig:1}b), which can lead to instability in certain
parameter regimes where the dynamics is otherwise stable when there
is no explicit PT-symmetry breaking by viscosity. 

The physics underpinning the instability triggered by viscosity through
explicit PT-symmetry breaking is that PT symmetry imposes strong constraints
on the dynamics of the system and only permits instability through
spontaneous PT-symmetry breaking. When these dynamic constraints are
removed by viscosity or other dissipative mechanism, the system becomes
less ``rigid'' or more ``slippery'', which can lead to instability
in certain parameter regimes that are stable for conservative systems. 

This interesting, if not surprising, effect of viscosity can be contrasted
with the viscosity induced total variation non-increasing in simple
systems, such as those governed by a scalar hyperbolic conservation
law. The main difference is that the dynamics of two-stream interaction
is more complex--containing more degrees of freedom. Dissipative
effects, including viscosity and friction, explicitly breaks the constraints
of PT-symmetry and exposes the dynamics to lower energy states, and
the process of jumping to lower energy states is accompanied by instabilities
and thus total variation growth. In comparison, simpler systems have
no such hidden lower energy state and viscosity only leads to total
variation non-increasing for field variables. 

We start from the following one dimensional two-fluid equation system
with viscosity,
\begin{alignat}{1}
\frac{\partial n_{j}}{\partial t}+\frac{\text{\ensuremath{\partial}}}{\partial z}\left(n_{j}v_{j}\right) & =0\thinspace,\label{eq:continuity}\\
\frac{\partial v_{j}}{\partial t}+v_{j}\frac{\text{\ensuremath{\partial}\ensuremath{v_{j}}}}{\partial z}-\nu_{j}\frac{\text{\ensuremath{\partial^{2}}\ensuremath{v_{j}}}}{\partial z^{2}}-\frac{q_{j}}{m_{j}}E+\frac{q_{j}}{n_{j}m_{j}}\frac{\text{\ensuremath{\partial}\ensuremath{p_{j}}}}{\partial z} & =0\thinspace,\label{eq:momentum}\\
\frac{d}{dt}\left(\frac{p_{j}}{\left(m_{j}n_{j}\right)^{\gamma_{j}}}\right) & =0\thinspace,\label{eq:energy}\\
\frac{\text{\ensuremath{\partial}}E}{\partial z}-\sum_{j}4\pi q_{j}n_{j} & =0\thinspace.\label{eq:poisson}
\end{alignat}
Here, $j=1,2$ is the index for species, and $q_{j},$ $m_{j,},$
$\gamma_{j}$, and $\nu_{j}$ are charge, mass, polytropic index,
and viscosity for each species. Equations (\ref{eq:continuity})-(\ref{eq:energy})
govern the dynamics of density, velocity, and pressure for each species.
The coupling between the two species is mediated by the electrostatic
field self-consistently determined by the Poisson equation (\ref{eq:energy}).
The equilibrium $(n_{j0},v_{j0},p_{j0},E_{0})$ is assumed to be homogeneous
with differential equilibrium flow for both species in the $z$-direction,
i.e., $v_{10}\ne v_{20}.$ There is no equilibrium electric field,
i.e., $E_{0}=0$ with $\sum_{j}q_{j}n_{j0}=0$. Consider a small perturbation
of the density, velocity, pressure, and electrostatic field of the
form
\begin{equation}
\tilde{n}_{j},\tilde{v}_{j},\tilde{p}_{j},\tilde{E}\sim\exp(ikz-i\omega t).
\end{equation}
The linear dynamics of the system can be cast into the form of Schrödinger's
equation,
\begin{align}
H\psi & =\omega\psi\thinspace,\label{Hpsi}\\
H & =\left(\begin{array}{cccc}
v_{10} & 0 & 1 & 0\\
0 & v_{20} & 0 & 1\\
\omega_{1}^{2}+v_{t1}^{2} & -\omega_{1}^{2} & v_{10}-i\nu_{1} & 0\\
-\omega_{2}^{2} & \omega_{2}^{2}+v_{t2}^{2} & 0 & v_{10}-i\nu_{2}
\end{array}\right)\thinspace,\label{H}\\
\psi & =\left(\tilde{n}_{1},\tilde{n}_{2},\tilde{v}_{1},\tilde{v}_{2}\right)^{T}\thinspace.\label{psi}
\end{align}
In deriving Eqs.\,(\ref{Hpsi})-(\ref{H}), the Poisson equation
(\ref{eq:poisson}) and energy equation (\ref{eq:energy}) have been
used as constraints to eliminate $\tilde{E}$ and $\tilde{p}_{j}$
in favor of $\tilde{n}_{j}.$ All quantities are dimensionless. A
characteristic velocity $V$ has been chosen to normalized all velocity
fields, and all frequencies variables, including $H$, are normalized
by $kV$. The density perturbation $\tilde{n}_{j}$ is normalized
by $n_{j0}$, and $\nu_{j}$ is normalized by $k/V$, $\omega_{j}^{2}=4\pi n_{j0}q_{j}^{2}/m_{j}k^{2}V^{2}$
is the normalized plasmas frequency squared, and $v_{tj}^{2}\equiv\gamma_{j}p_{j0}/m_{j}V^{2}$
is the normalized thermal velocity squared. The system is determined
by 8 dimensionless parameters: $(\nu_{1},\nu_{2},v_{t1},v_{t2},v_{10},v_{20},\omega_{1},\omega_{2}).$

Equation (\ref{Hpsi}) assumes the form of Schrödinger's equation,
but the Hamiltonian $H$ is certainly not Hermitian, which is typical
in classical systems. One of the characteristics of non-Hermitian
systems, such as symplectic \citep{Arnold89-all,deGosson06}, pseudo-Hamiltonian
\citep{Dirac1942,Pauli1943,Krein1950,Gelfand55,Lee1969,Yakubovich75},
and PT symmetric \citep{Bender1998,Bender2007,Bender2007a,Bender2010,Felski2021,Bender2014,Zhang2020PT}
systems, is that they admit instabilities, i.e., eigen frequency $\omega$
with $\text{Im\ensuremath{(\omega)>0}}.$ The dynamics of non-Hermtian
systems is more complex. For the two-stream interaction, charting
the stability diagram in the 8D parameter space can be challenging.
A systematic approach for analyzing the parameter dependence of instabilities
is afforded by PT-symmetry analysis \citep{Qin2019KH,Zhang2020PT,Fu2020KH,Qin2021}.
Here we briefly summarize this method. 

For a non-Hermitian Hamiltonian $H,$ we may ask if it respects other
symmetries of physical importance. One such symmetry is PT symmetry,
originated from the Lorentz group, the homogeneous symmetry of flat
spacetime. The $PT$ transformation is a discrete element of the Lorentz
group $O(1,3),$ which as a topological manifold contains 4 disconnected
components. In general, we expect that physics is invariant with respect
to only one of the components--the proper orthochronous Lorentz group
$SO^{+}(1,3).$ The whole Lorentz group $O(1,3)$ is a semi-direct
product of $SO^{+}(1,3)$ and the discrete subgroup $\{1,\mathcal{P},\mathcal{T},\mathcal{P}\mathcal{T}\}$,
i.e., 
\begin{equation}
O(1,3)=SO^{+}(1,3)\rtimes\{1,\mathcal{P},\mathcal{T},\mathcal{P}\mathcal{T}\},
\end{equation}
where $\mathcal{P}=\mathrm{diag}(1,-1,-1,-1)$ and $\mathcal{T}=\mathrm{diag}(-1,1,1,1).$
Knowing that physics is not invariant with respect to $\mathcal{P}$
transformation or $\mathcal{T}$ transformation, we can ask a weaker
question: Is physics invariant with respect to the $\mathcal{P}\mathcal{T}$
transformation \citep{BenderPrivate2019}? Bender initiated the program
to study the physics associated with $\mathcal{P}\mathcal{T}$ transformation
\citep{Bender1998,Bender2007,Bender2007a,Bender2010,Felski2021,Bender2014,Zhang2020PT}.
It was found that for conservative classical systems, including those
in neutral fluids and plasmas, PT symmetry can result from reversibility
\citep{Qin2019KH,Zhang2020PT,Fu2020KH,Qin2021}. This finding is consistent
with Bender's original characterization of PT-symmetry as a loss-grain
balance between two coupled subsystems. A Hamiltonian $H$ is called
PT symmetric if it commute with a PT operator, 
\begin{equation}
PTH-HPT=0,
\end{equation}
where $T$ denote complex conjugate and $P$ is a parity operator,
i.e., $P^{2}=I.$ Note that here we have generalized the $\mathcal{P}$
element of the Lorentz group to an arbitrary parity operator. A PT
symmetric $H$ must has a spectrum that is symmetric with respect
to the the real axis, i.e., for every unstable mode $\omega_{R}+i\omega_{I}$
($\omega_{I}>0$), there must exist a damped mode $\omega_{R}-i\omega_{I}$.
This topological constraint in spectrum space stipulates that in order
for a stable PT symmetric system to become unstable when the system
parameters vary, the stable eigenmodes on the real axis must collide
(resonate) first. It was recently proved \citep{Zhang2020PT} that
this collision is precisely the Krein collision \citep{Krein1950,Gelfand55,Yakubovich75}
in pseudo-Hermitian systems. This process is illustrated in Fig.\,\ref{fig:1}a,
and it is the only route for a PT symmetric system to become unstable.
Krein collision is a necessary but not sufficient condition for instability.
Only collision between a negative-action mode (denoted by blue in
Fig.\,\ref{fig:1}a) and a positive-action mode (denoted by red)
will lead to instability. A stable eigenmode itself is invariant under
PT transformation, but an unstable or damped eigenmode is not, even
though the governing equations are always PT symmetric. Thus, the
destabilization of the system via Krein collisions breaks PT symmetry
spontaneously. Without considering their geometric properties, the
locations in the parameter space where instabilities are triggered
were previously called exception points or thresholds. In a PT symmetric
system, a stable mode is not allowed to wonder off from the real axis
to become unstable by itself. The route of destabilization illustrated
in Fig.\,\ref{fig:1}b is forbidden by PT symmetry. 

However, if the PT-symmetry condition is explicitly broken, for example
by a dissipative mechanism of viscosity or friction, then the constraint
requiring spectrum being symmetric with respect to the real axis is
removed, and the route to instability in Fig.\,\ref{fig:1}b forbidden
by PT-symmetry is permitted. This is the mechanism of dissipation
induced instability via explicit PT-symmetry breaking. 

Instability driven by dissipation is an intriguing phenomena \citep{Kirillov2013,Kirillov2013a}.
In plasma physics, resistivity induced instabilities, including the
tearing mode and the resistive-wall mode, belong to this category.
Nevertheless, there exists an obvious question or doubt here: Dissipation
always takes energy away from the system by definition; How can it
drive the system more unstable? This issue has not been satisfactorily
addressed in the literature. The physical picture presented here emphasizes
the role of explicit PT-symmetry breaking by dissipation. A conservative
system may possess lower energy states, but they are not accessible
due to the PT-symmetry constraints. Dissipation explicitly breaks
these constraints and expose the system to the lower energy states,
albeit it may consume energy to do so. When the exposed energy drop
dominates the dissipation, the system is destabilized. 

\begin{figure}[ht]
\centering

\includegraphics[width=7cm]{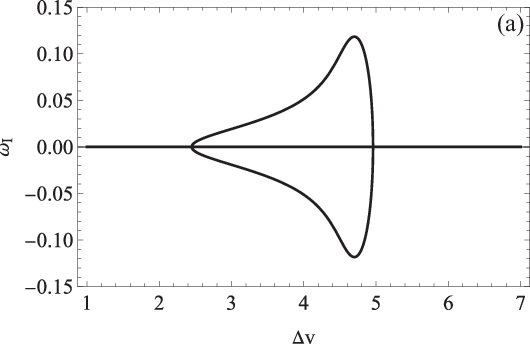} \includegraphics[width=7cm]{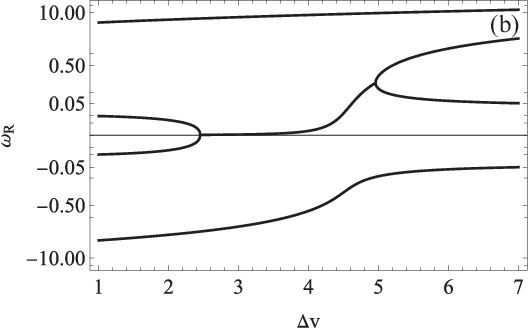} 

\includegraphics[width=8cm]{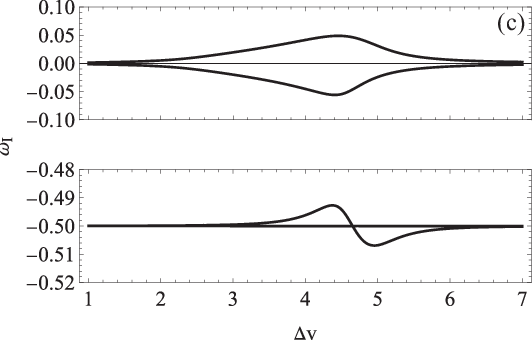} \includegraphics[width=7cm]{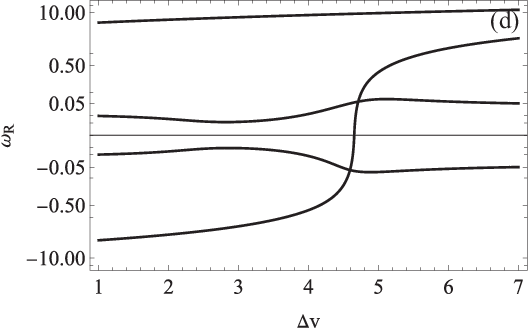}\caption{Two-stream instabilities for $\omega_{2}/\omega_{1}=100.$ Plotted
are the four eigen-frequencies $\omega_{R}+i\omega_{I}$ of the system
as functions of the differential velocity $\Delta v$. Absolute log
scale was used for (b) and (d). The large empty space between $-1.48<\omega_{I}<-0.10$
is omitted in (c) for better resolution. The system parameters for
(a) and (b) are $(\nu_{1},\nu_{2},v_{t1},v_{t2},v_{10},v_{20},\omega_{1},\omega_{2})=(0,0,0,6,0,\Delta v,0.04,4).$
There are two Krein collisions in (a) at $\Delta v=2.44$ and $\Delta v=4.96$.
The system parameters for (c) and (d) are $(\nu_{1},\nu_{2},v_{t1},v_{t2},v_{10},v_{20},\omega_{1},\omega_{2})=(0,1,0,6,0,\Delta v,0.04,4).$ }
\label{fig:2}
\end{figure}

\begin{figure}[ht]
\centering

\includegraphics[width=7cm]{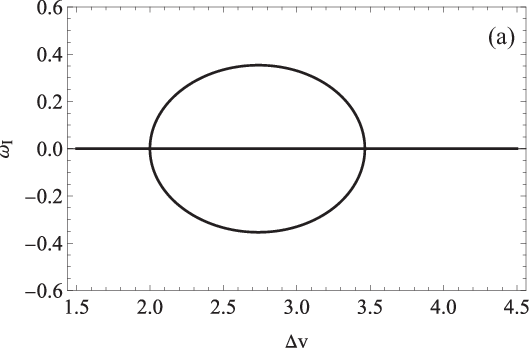} \includegraphics[width=7cm]{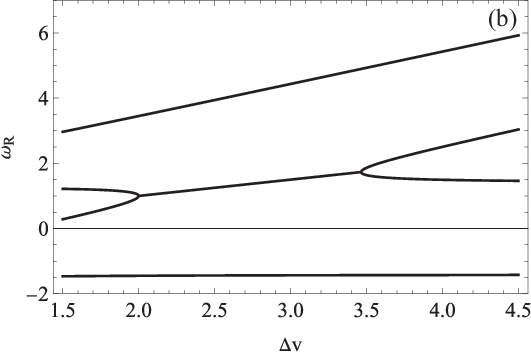}

\includegraphics[width=7cm]{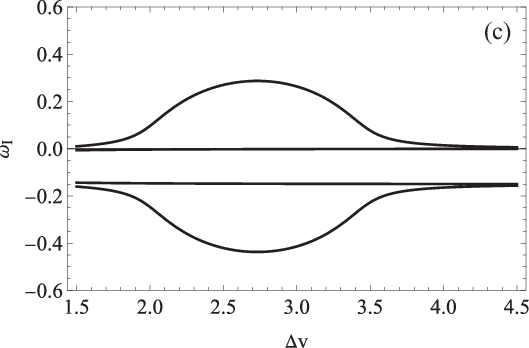} \includegraphics[width=7cm]{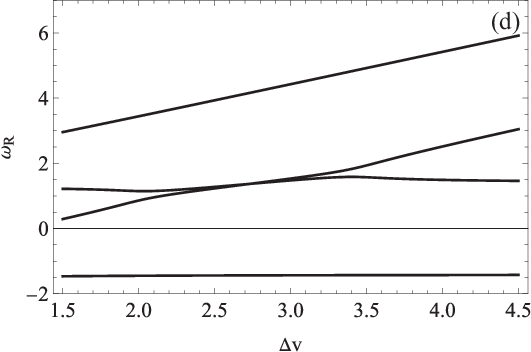}\caption{Two-stream instabilities for $\omega_{2}/\omega_{1}=1.$ Plotted are
the four eigen-frequencies $\omega_{R}+i\omega_{I}$ of the system
as functions of the differential velocity $\Delta v$. The system
parameters for (a) and (b) are $(\nu_{1},\nu_{2},v_{t1},v_{t2},v_{10},v_{20},\omega_{1},\omega_{2})=(0,0,1,1,0,\Delta v,1,1).$
There are two Krein collisions in (a) at $\Delta v=2$ and $\Delta v=3.46$.
The system parameters for (c) and (d) are $(\nu_{1},\nu_{2},v_{t1},v_{t2},v_{10},v_{20},\omega_{1},\omega_{2})=(0,0.3,1,1,0,\Delta v,1,1).$ }
\label{fig:3}
\end{figure}

Going back to the Hamiltonian specified by Eq.\,(\ref{H}) for the
two-stream interaction. When the viscosity $\nu_{1}$ and $\nu_{2}$
vanish, the system is trivially PT symmetric, for $P=I.$ The spectrum
is determined by the dispersion relation
\begin{equation}
\frac{\omega_{1}^{2}}{\left(\omega-v_{10}\right)^{2}-v_{t1}^{2}}+\frac{\omega_{2}^{2}}{\left(\omega-v_{20}\right)^{2}-v_{t2}^{2}}=1\thinspace.\label{DR}
\end{equation}
As discussed above, the only route to instability is through the Krein
collision between a positive-action mode and a negative-action mode
\citep{Zhang2016GH}. When the instability is triggered, PT symmetry
is spontaneously broken. Two numerically calculated examples are displayed
in Figs.\,\ref{fig:2}a and \ref{fig:2}b for the case of mass ratio
between two charge components being 100, and in Figs.\,\ref{fig:3}a
and \ref{fig:3}b for the case of mass ratio being 1. Plotted in Figs.\,\ref{fig:2}a
and \ref{fig:3}a are the imaginary parts of the 4 eigen-frequencies
as functions of the relative equilibrium velocity $\Delta v\equiv v_{20}-v_{10}.$
Figures \ref{fig:2}b and \ref{fig:3}b are the real parts of the
eigen-frequencies. The system parameters in Figs.\,\ref{fig:2}a
and \ref{fig:2}b are $(\nu_{1},\nu_{2},v_{t1},v_{t2},v_{10},v_{20},\omega_{1},\omega_{2})=(0,0,0,6,0,\Delta v,0.04,4).$
There are two Krein collisions in Fig.\,\ref{fig:2}a at $\Delta v=2.44$
and $\Delta v=4.96$, and the system is unstable for $2.44<\Delta v<4.96.$
The system parameters in Figs.\,\ref{fig:3}a and \ref{fig:3}b are
$(\nu_{1},\nu_{2},v_{t1},v_{t2},v_{10},v_{20},\omega_{1},\omega_{2})=(0,0,1,1,0,\Delta v,1,1).$
There are two Krein collisions in Fig.\,\ref{fig:3}a at $\Delta v=2$
and $\Delta v=3.46$, and the system is unstable for $2<\Delta v<3.46.$
Observe that the spectrum in Figs.\,\ref{fig:2}a and \ref{fig:3}a
is always symmetric with respect to the real axis, as required by
PT symmetry.

For the $H$ defined in Eq.\,(\ref{H}), when the viscosity is non-vanishing,
PT symmetry of the system is explicitly broken. It can be proven that
there exist no parity operator $P$ such that $PTH=HPT.$ Such cases
are displayed in Figs.\,\ref{fig:2}c, \ref{fig:2}d, \ref{fig:3}c,
and \ref{fig:3}d. The system parameters for Figs.\,\ref{fig:2}c
and \ref{fig:2}d are $(\nu_{1},\nu_{2},v_{t1},v_{t2},v_{10},v_{20},\omega_{1},\omega_{2})=(0,1,0,6,0,\Delta v,0.04,4),$
which are identical to those for Figs.\,\ref{fig:2}a and \ref{fig:2}b
except for the non-vanishing $\nu_{2}$. Because the governing system
is not PT symmetric anymore, the spectrum is not symmetric with respect
to the real axis and the route to instability illustrated in Fig.\,\ref{fig:1}b
is now possible. The system is unstable for entire plotted interval
of $1<\Delta v<7$. In particular, the intervals $1<\Delta v<2.44$
and $4.96<\Delta v<7$ are stable regions when the viscosity vanishes.
The two-stream instability in these two regions are triggered by the
explicit PT-symmetry breaking induced by finite viscosity. A similar
situation is found in Figs.\,\ref{fig:3}c and \ref{fig:3}d, where
the system parameters are $(\nu_{1},\nu_{2},v_{t1},v_{t2},v_{10},v_{20},\omega_{1},\omega_{2})=(0,0.3,1,1,0,\Delta v,1,1),$
which are the same as those for Figs.\,\ref{fig:3}a and \ref{fig:3}b
except for the non-vanishing $\nu_{2}$. Now the system is unstable
for the entire plotted interval of $1.5<\Delta v<4.5$, which includes
two intervals, $1.5<\Delta v<2$ and $3.46<\Delta v<7$, that are
stable regions when there is no viscosity. 

As discussed above, viscosity induced instability reveals that in
addition to being an dissipative effect, viscosity can also reduce
the rigidity of the dynamics and consequently inject free energy into
the system. In the current context, the rigidity is represented by
the PT-symmetry constraints of the conservative system. In general,
complex system with large degrees of freedom is more likely to have
this type of hidden free energy protected by the PT-symmetry constraints.
The two-stream interaction described by Eqs.\,(\ref{eq:continuity})-(\ref{eq:poisson})
is such a system. Specifically, the free energy is hidden in the differential
equilibrium flow and pressure between the two components, as indicated
in Eq.\,(\ref{DR}). In comparison, a scalar field $u(t,z)$ governed
by a hyperbolic conservation law
\begin{equation}
\frac{\partial u}{\partial t}+\frac{\partial}{\partial z}f(u)=0\label{dudt}
\end{equation}
does not have this type of free energy accessible by viscosity. Here,
$f(u)$ is a given function of $u$. In fact, the following result
can rigorously proven \citep{Shu2006}: Let $\nu>0,$ and $u^{\nu}(t,z)$
is the solution of the following initial value problem, 
\begin{align}
\frac{\partial u^{\nu}}{\partial t}+\frac{\partial}{\partial z}f(u^{\nu}) & =\nu\frac{\partial^{2}u^{\nu}}{\partial z^{2}}\thinspace,\label{dundt}\\
u^{\nu}(t=0,z) & =u_{0}(z)\thinspace,
\end{align}
then the total variation of $u^{\nu}$ defined by
\begin{equation}
\text{TV\ensuremath{\left[u^{\nu}\right]}\ensuremath{\ensuremath{\equiv\sup_{h}\int\left|\frac{u(z+h)-u(z)}{h}\right|}dz}}
\end{equation}
is non-increasing with time. Assuming $u^{\nu}$ is smooth, we have
\begin{align}
\text{TV\ensuremath{\left[u^{\nu}\right]}\ensuremath{=}} & \int\left|\frac{\partial u}{\partial z}\right|dz\thinspace,\\
\frac{d}{dt}TV\left[u^{\nu}\right] & =\frac{d}{dt}\int\left|\frac{\partial u}{\partial z}\right|dz\le0\thinspace.
\end{align}

For the scalar hyperbolic conservation law (\ref{dudt}), the entropy
solution is defined to be $u(t,z)\equiv\lim_{\nu\rightarrow0}u^{\nu}(t,z)$.
When numerically solving Eq.\,(\ref{dudt}) for an entropy solution,
it is desirable to adopt an algorithm that preserves the property
of non-increasing total variation for the exact entropy solutions
\citep{Harten1983,Yee1985,Cockburn1989}. 

If one chooses to, Eq.\,(\ref{dundt}) can be viewed as an extremely
simplified version of the two-stream system described by Eqs.\,(\ref{eq:continuity})-(\ref{eq:poisson})
when the two components are cold $(p_{j}=0)$ and do not interact
$(E=0)$, which stripes away all possible hidden free energy and total
variation cannot grow. 

Obviously, the properties of non-increasing total variation at finite
viscosity does not hold for the general two-stream system governed
by Eqs.\,(\ref{eq:continuity})-(\ref{eq:poisson}). As we have shown,
in certain certain parameter regimes, the linear dynamics of the system,
which is otherwise stable, can be destabilized by finite viscosity
via explicit PT-symmetry breaking. For these cases, at the equilibrium
$(n_{10},n_{20},v_{10},v_{20},p_{10},p_{20}),$ when the initial perturbation
is small, the dynamics is dominated by the linear eigenmodes for a
short time at $0<t<\delta.$ If the initial perturbation takes the
form of the an unstable eigenmode with a given $k$ and $\omega=\omega_{R}+i\omega_{I}$
($\omega_{I}>0$), then we have
\begin{equation}
\frac{d}{dt}\int\left|\frac{\partial E}{\partial z}\right|dz\sim\omega_{I}e^{\omega_{I}t}\int\left|kE\right|dz>0\,,
\end{equation}
i.e., the total variation of the electric field $E$ increases with
time. The same result holds for all other field variable as well.
Of course, at zero viscosity, the two-stream interaction can be physically
unstable via spontaneous PT-symmetry breaking, resulting in total
variation growth as well. The complexity of the dynamics in plasmas
with different charge components needs to be considered when designing
or selecting algorithms for numerical solutions. 
\begin{acknowledgments}
This research was supported by the U.S. Department of Energy (DE-AC02-09CH11466).
\end{acknowledgments}

\bibliographystyle{apsrev4-2}
\bibliography{D:/W/Refs/Refs}

\end{document}